# High-throughput in-situ characterization and modelling of precipitation kinetics in compositionally graded alloys

F. De Geuser[1,2], M.J. Styles[3], C.R. Hutchinson[4] and A. Deschamps[1,2]

[1]Univ. Grenoble Alpes, SIMAP, F-38000 Grenoble, France

[2]CNRS, SIMAP, F-38000 Grenoble, France

[3]CSIRO Manufacturing Flagship, Clayton, 3168, VIC, Australia

[4]Department of Materials Science and Engineering, Monash University, Clayton, 3800, VIC, Australia

## Abstract

The development of new engineering alloy chemistries is a time consuming and iterative process. A necessary step is characterization of the nano/microstructure to provide a link between the processing and properties of each alloy chemistry considered. One approach to accelerate the identification of optimal chemistries is to use samples containing a gradient in composition, ie. combinatorial samples, and to investigate many different chemistries at the same time. However, for engineering alloys, the final properties depend not only on chemistry but also on the path of microstructure development which necessitates characterization of microstructure evolution for each chemistry.In this contribution we demonstrate an approach that allows for the in-situ, nanoscale characterization of the precipitate structures in alloys, as a function of aging time, in combinatorial samples containing a composition gradient. The approach uses small angle x-ray scattering (SAXS) at a synchrotron beamline. The Cu-Co system is used for the proof-of-concept and the combinatorial samples prepared contain a gradient in Co from 0% to 2%. These samples are aged at temperatures between 450ºC and 550ºC and the precipitate structures (precipitate size, volume fraction and number density) all along the composition gradient are simultaneously monitored as a function of time. This large dataset is used to test the applicability and robustness of a conventional class model for precipitation that considers concurrent nucleation, growth and coarsening and the ability of the model to describe such a large dataset.

# 1  Introduction

The process of designing engineering alloys requires finding an optimal point in a chemistry space, subject to relevant constraints of processing. In all major alloy families, the number of alloying elements typically ranges from 5 to 10 and optimizing a composition in such multicomponent space is an extremely challenging task. The traditional approach is to try to decouple the interactions between different solutes and study alloys of discrete compositions. In recent years, considerable efforts have been launched (e.g. the Materials Genome Initiative (MGI)[1] and the Accelerated Metallurgy Project[2]) to develop new alloy design strategies using combinatorial methods in both computational material science and related experimental approaches that accelerate this alloy design process. Materials containing compositional gradients have long been used to map the composition space of alloys. Specifically, diffusion couples or multiples have been used for determining the effect of composition on the material structure (phase diagram identification[3]–[7], on the diffusion of solute species[8], and on various properties related to materials chemistry (e.g. modulus[9], thermal conductivity[10], [11]) and sometimes more complex properties such as shape memory alloy identification [12] or metallic glass formability [13].

However, a key characteristic of engineering metallic alloys is that their main properties depend not only on chemistry, but also on microstructure. Therefore, their optimization with respect to a given property requires an understanding of the effect of chemistry on the kinetic path of microstructure development during thermal or thermo-mechanical treatments. Few studies have actually tried to determine the microstructure development in compositionally graded materials from the point of view of combinatorial experimentation. The community interested in phase transformations in steels have used this approach to study the compositional limits for certain types of phase transformations[14], [15]: the limits for the massive transformation[16], [17], acicular ferrite formation[18] and allotriomorphic ferrite formation[19], [20] have all been examined using specially designed samples containing a macroscopic gradient in either carbon or substitutional solute such as Ni or Mn. Sinclair et al. [21] have even used a sample containing a gradient in Nb content in one direction and a gradient in temperature in an orthogonal direction to simultaneously probe the effect of temperature and Nb content on the recrystallization of Fe. Similar approaches have been used by to study the compositional limits for coherent versus incoherent precipitation [22], order/disorder transformations[22], [23], competition between spinodal decomposition and nucleation and growth[22], nucleation in the vicinity of phase boundaries [22], [24]and the effect of Cu and Mg content on the rapid hardening phenomenon in Al-Mg-Cu alloys [25].

In all of these studies, the alloys were observed ex-situ after a specific heat treatment, which provided a snapshot of the microstructure (and hence of the corresponding potential properties) as a function of chemical composition. However, composition interacts in a complex manner with microstructure development during heat treatments, and an alloy design strategy requires the characterization of the full kinetic path in composition space. If this is to be done, in a combinatorial manner, by exploiting samples containing a macroscopic composition gradient, the characterization strategy requires tools exhibiting the following characteristics:

- Fast, quantitative characterization of the microstructural feature of interest.
- Spatially resolved with a high resolution compared to the size of the composition gradient, yet probing a volume large enough to guaranteesufficiently good statistics on the measured microstructure features.
- Time resolved with a time resolution sufficient so the kinetics can be monitored simultaneously in the desired number of locations within the composition gradient.

For the particular case of strengthening precipitation, the microstructural features of interest are the size, volume fraction and number density of precipitates. In most metallic systems the precipitate radius for maximum strength occurs at the nanoscale (often 2-5nm). The only experimental technique that satisfies the three conditions above is Small-Angle X-ray Scattering (SAXS). Reviews discussing the manner in which it can be used to provide a fast, quantitative characterization of precipitates can be found in[26]–[30]. The spatial resolution of these measurements is equal to the X-ray beam size, which is typically greater than 1 mm for laboratory sources and 100 μm for synchrotron experiments. Several studies have demonstrated the capability of SAXS to map the distribution of nanoscale precipitates in heterogeneous microstructures such as in welds [31]–[33]. Furthermore, SAXS is particularly well suited to be performed in-situ during heat treatments, along isothermal or more complex thermal paths (e.g.[34], [35]). Depending on the particular contrast of scattering factors between the precipitates and matrix, acquisition times on synchrotron beamlines can be as low as 1-10s, which opens the possibility to couple spatially and time-resolved experiments.

The aim of this contribution is to demonstrate the feasibility of in-situ, combinatorial studies of the effect of alloy composition on precipitation kinetics, and to use the acquired database as a tool for assessing the capability of a simple precipitation model in a wide range of compositions and temperatures. For the experimental proof-of-concept study, we have chosen a model system, Cu-Co. This system offers a number of advantages, including:

- This system is relatively dilute (a maximum of 2wt%Co is used), and the precipitates formed are almost pure Co[36]. These two conditions will make it possible to apply relatively simple precipitation models.
- At relatively low temperatures, the precipitates form as spherical particles[37]–[39], which simplifies the interpretation of the SAXS data;
- Cu-Co has been used several times as a model system (ie. spherical precipitates of pure Co exhibiting negligible strain with the matrix) to assess the capability of classical nucleation theory and more generally to provide a comparison with precipitation models [40], [41].

As will be detailed below, the experiments involved preparing diffusion couples between pure Cu and Cu-2wt%Co, performing a solution heat treatment on the composition gradient material, and subsequently performing heat treatments at three temperatures (450, 500 and 550°C) in-situ at a synchrotron SAXS beamline (BM02 – D2AM at ESRF) while measuring the SAXS signal at different positions along the composition gradient (Fig. 1). Co undergoes an allotropic phase transformation at 422°C which is below the lowest temperature studied in this work. As a result Co precipitates as FCC particles under all conditions considered in this contribution.Experimental difficulties were yet encountered. They included some loss of Co during diffusion couple preparation that must be accounted for. The grain structure within the sample was such that double Bragg scattering interfered with the measured SAXS signal, and it was necessary to develop a specific methodology to deconvolute the two signals. Finally, a challenge was to link the observed precipitation kinetics to the local alloy composition where the X-ray measurement was made. A specific procedure was used to obtain an in-situ composition measurement with the help of the X-ray beam.

The precipitation kinetics, as a function of time, temperature and solute content has been modelled using the so-called numerical Kampmann-Wagner class model [42], [43]. In relatively simple systems, such as that studied here, this model has been shown to provide a robust and efficient frameworkwhich compares satisfactorily with experimental results [43]–[46]. One advantage of this modelling technique is its good computational efficiency, which makes it possible to apply to a large set of experimental conditions with varying temperature and alloy composition, so that theresults can be compared to the full dataset that is produced from the combinatorial experiment shown in Fig. 1. One of the interests here is to assess the limitations (and therefore the robustness) of this model when applied to situations where all parameters (e.g. diffusivity, driving force) vary widely.

## 2 Materials and preparation of the diffusion couple

The diffusion couples were prepared from pieces of Pure Cu and Cu-2.0Co (wt. %) produced by AMES Laboratory and contained impurity levels below 0.01%, as measured by ICP-AES. Pieces of pure Cu and Cu-2Co were first sectioned into cubes ~14mm x 14mm x 14mm. To help the bonding between the alloys, pieces of each alloy were held together for ~10min at ~500ºC using a hot compression machine before being encapsulated in a quartz tube and transferred to a tube furnace for 15 days at 1000ºC (above the solvus boundary for Cu-2Co). Diffusion calculations suggest that the Co gradient would be 500-600 μm after such a heat treatment. To further spread the compositional gradient, each diffusion couple was subsequently hot compressed in a channel die that constrained the deformation to the direction of the diffusion zone. This channel die hot compression was performed at 700ºC with a strain rate of $2 \times 10^{-3}$ $s^{-1}$. The samples were reduced in height from ~15mm to 7mm which approximately doubled the length of the diffusion zone.

Samples were then solution treated at 1000ºC for 1h, water quenched and sectioned into thin slices using a precision saw. The sample slices were subsequently thinned to ~30-50 μm using grinding paper ready for in-situ examination using small angle x-ray scattering.

## 3 Characterization of the composition gradient

Due to the small size of the diffusion zone in the samples (1-3mm), and the uncertainty regarding the alignment of the samples relative to the X-ray beam, it is challenging to ascertain precisely at which position in the composition gradient the X-ray beam is aiming. It is therefore desirable to use the beam itself to measure the local concentration of Co in the samples after they have been mounted in the furnace. This can be achieved by monitoring the change in transmitted beam intensity as a function of distance along the diffusion couple at two beam energies, above and below the Co K absorption edge (7.709 keV). The relationship between the local transmission and the properties of the sample can be expressed as:

$$T_x = e^{-\mu_x t_x} \qquad (1)$$

where $T_x$ is the local transmission, and $\mu_x$ and $t_x$ are the linear absorption coefficient and the thickness of the foil, respectively, at position $x$ on the diffusion couple. On the BM02 - D2AM beamline, the transmission of the sample is measured by comparing the ratio of the upstream and downstream beam monitors with and without sample (at each position *x*).The local thickness of the sample can be expressed as:

$$t_x = -\frac{\ln(T_x)}{\mu_x} \quad (2)$$

The sample thickness is independent of the beam energy, while both $\mu_x$ and $T_x$ depend on the energy. If the transmission measurement is performed at two energies, below and above the absorption edge (at 7.7 and 7.8 keV), the variation of the local transmission can be used to deduce how the local absorption coefficient $\mu_x$ varies at both energies. This variation can then be linked to the local Co concentration:

$$\frac{\mu_{x\_7.7keV}}{\mu_{x\_7.8keV}} = \frac{\ln(T_{x\_7.7keV})}{\ln(T_{x\_7.8keV})} = A$$

$$\frac{\mu_{x\_7.7keV}}{\mu_{x\_7.8keV}} = \frac{(1-c_x)\mu_{Cu\_7.7keV} + c_x\mu_{Co\_7.7keV}}{(1-c_x)\mu_{Cu\_7.8keV} + c_x\mu_{Co\_7.8keV}}$$

$$c_x = \frac{\mu_{Cu\_7.7keV} - A\mu_{Cu\_7.8keV}}{\mu_{Cu\_7.7keV} - \mu_{Co\_7.7keV} + A(\mu_{Co\_7.8keV} - \mu_{Cu\_7.8keV})} \quad (3)$$

where $c_x$ is the local concentration (at.%) of Co, and $\mu_{Cu}$ and $\mu_{Co}$ are the linear absorption coefficients for pure Cu and pure Co, respectively, calculated for the two different beam energies using the density of Cu (8.96 g/cm$^3$). Once $c_x$ has been obtained, $\mu_x$ and $t_x$ can be calculated. The results of these calculations are shown in Fig. 2 for the three diffusion couples. Fig.2a confirms that the method is able to detect the local Co content (more detailed discussion on the actual profile follows further in the paper). Each diffusion couple is shown to have a constant (or slowly varying) thickness along the probed direction despite the varying Co content (Fig. 2b). This is an *a posteriori* confirmation that the local content has been properly taken into account by this approach.

In order to corroborate the measurements of the local alloy composition made during the in situ SAXS experiment, EPMA measurements were performed on the foils following the heat treatment. Backscattered electron (BSE) micrographs collected from a solution treated and quenched diffusion couple (i.e. the starting material for the in-situ SAXS experiments) are shown in Figure 3. While the two ends of the diffusion couples were found to be homogeneous, these images revealed the presence of Co oxide inclusions situated near the middle of the diffusion zone. It is thought that these inclusions were introduced at the interface between the two starting alloys during the joining process, and subsequently distributed through the diffusion zone during the heat treatment and channel die compression. Similar inclusions were found in all three of the samples aged in-situ. These inclusions have the effect of removing Co from solid solution and this must be taken into account during the modelling of the kinetics of the precipitation process.

Quantitative measurements of the Co concentration profile were performed using EPMA and the results are shown in Figure 4. As expected, no Co was detected in the pure Cu end. An average of 1.82 at.% of Co was detected in the Co rich end, which is slightly lower than the expected 2.03 at.% measured by ICP chemical analysis of the Cu-2wt%Co alloy from which the couple was prepared. Over the diffusion zone, the Co concentration increased from 0 to 1.82 at.% over the space of ~3 mm in this sample, and followed an approximately sigmoidal profile, also shown in Figure 4.

Equivalent analyses were performed on the Cu-Co diffusion couples aged in-situ, and the results are compared with the SAXS measurements in Figure 5. It can be seen that the profile of the SAXS results are in reasonable agreement with the profiles measured by EPMA, however, the maximum concentrations determined by EPMA are typically lower than the SAXS results. This is consistent with the hypothesis that the oxides particles consume a significant amount of Co, lowering the Co available in solid solution. The SAXS beam size was ~200 μm in diameter, and passes through the whole foil (30-50 μm), so that the Co oxide particles are included in the analysis, whereas the spot size for the EPMA measurements was ~2 μm and only occasionally intersects the oxides. The insets from Figure 5 show the linear relationship between the EPMA and the SAXS results. From this comparison, it is reasonable to assume that the EPMA results are more representative of the Co solid solution composition available for precipitation. Without the combination of the two techniques, it would have been very difficult to ascertain the position of the beam in the composition gradient.

# 4  SAXS data processing

All of the SAXS images collected in this investigation contained streaks (e.g. Fig.6a) which varied in number and intensity with the sample position. Generally, the number of streaks appeared to increase towards the pure Cu end of the diffusion couples. This observation, combined with the high purity and large grain size of the Cu foils, suggests that the streaks are multiple-diffraction effects. As the samples were heated in-situ, the position and intensity of the streaks were observed to evolve as the result of the slight displacement of the sample due to thermal expansion. However, once the aging temperature was reached, the streaks at each sample position were found to be constant over time.

The intensity of the streaks, particularly near the beam-stop (low q region), often obscured the underlying SAXS signal, and prevented straightforward azimuthal integration of the data. A total masking of the streaks in the images would remove too large an area from the integration, resulting in poor statistics and a possibly limited q range (especially in the low q region). Since it was observed that the streaks were stable after reaching the aging temperature, it was decided to subtract the initial image of each position on the diffusion couple from the subsequent images. While this does not completely remove the streaks (cf. Figure 6), it greatly improves the isotropy of the small angle scattering expected from the spherical precipitates. Since the initial image is used as a "background" for subsequent images, only variation from this initial stage is thus recorded. The obtained "streak-corrected" images are then azimuthally averaged. This procedure leads to zero scattering for the first image. While the amount of precipitates is expected to be small initially (the initial image is always recorded within 10 min after the sample reached aging temperature), it is likely to be non-zero. To account for this, we added to all the spectra an initial azimuthal integration of the first image, obtained with manually placed masks on the streaks. Overall, the manual masking resulting in a relatively lower signal over noise ratio applies only on the initial scattering, which is expected to be very small since the precipitation reaction has just commenced.

## 5  Precipitation kinetics: in situ SAXS measurements

The SAXS patterns were fitted to the theoretical scattering of an assembly of spherical precipitates of concentration 90% Co, which is close to experimental reports[36]. To account for the Laue scattering of the solid solution as well as for the contribution of the remaining streak features on the scattering pattern, we considered a background contribution of the form[47], [48]:

$$I_{Bg} = \frac{A}{q^n} + B$$

with the Porod exponent $n$ being typically between 3.5 and 4. The total intensity for a given SAXS pattern is then written[49]:

$$I = \Delta\rho^2 \int_0^\infty f(r) I_{sph}(q,r) dr + I_{Bg} \qquad (4)$$

$\Delta\rho^2$ is the scattering contrast between matrix and precipitates. It varies slightly with the position since it depends on the matrix Co content. $I_{sph}(q,r)$ is the intensity scattered by a single sphere of radius $r$. $f(r)$ is the size distribution function. We chose to use a Schulz distribution[50]. As is the case for other commonly used distributions (e.g. Gaussian, Lorentzian, log-normal…), it can be characterized by only two parameters, an average radius *Rm* and a polydispersity factor. For realistic polydispersities observed in engineering alloys, e.g.[51], [52], the Schulz distribution is close to a log-normal distribution. Its main advantage, however, is that with a Schulz distribution, Eq.(4)yields an analytical solution [53], so that numerical integration is not required and parameter fitting is much faster. This is particularly useful for time and space resolved experiments where the number of SAXS patterns to fit can be large.

A subset of the SAXS pattern is shown in Figure 7 for a single time (~10h) at the 3 studied temperatures 550°C (Fig. 7a), 500°C (Fig. 7b) and 450°C (Fig. 7c). The fitted intensities are also shown on the figure. The colour code used for the plot is related to the local Co content (lighter: less Co; darker: more Co). The samples generally show a higher intensity (i.e. a higher volume fraction) for higher local Co contents (darker colours). This intensity is mostly originated from a q-range which is lower for higher temperature, and higher for lower temperature. This agrees with the expected trend of forming smaller precipitates at lower temperature.

Figure 8 shows the SAXS results for the complete kinetics along the whole Co gradient for the 3 temperatures (450°C, 500°C, 550°C). It represents the evolution of the average size, the volume fraction and the number density of precipitates. Again, a qualitative general consistency can be observed. For a given temperature, the higher Co content gives rise to a higher volume fraction. Together with slower kinetics, a lower temperature induces smaller precipitates. The experimental data also indicates that a pure nucleation regime (i.e. constant size with increasing number density), if present, is out of reach of this experimental setup given the temporal resolution.

## 6  Precipitation kinetics: modelling

This large experimental data set can now be compared to a precipitation model. It represents a challenge for the robustness of such a model, since it covers the complete precipitation kinetics (nucleation, growth and coarsening) at 3 different temperatures over a significant concentration range.A classical numerical Kampmann-Wagner class model approach was used, the details of which can be found in [43]. We simply recall here that the model uses the classical equations for homogeneous nucleation, growth and coarsening of precipitates to simulate the evolution of an assembly of precipitates and monitors the full precipitate size distribution rather than a single mean precipitate size. The general principle is that each time step where nucleation occurs populates a size class. This size class subsequently evolves following the classical spherical growth equation, assuming local equilibrium at the interface. Because of interfacial energy, the equilibrium concentration at the interface is size dependent following the Gibbs-Thomson equation.

For a binary alloy giving rise to homogeneous precipitation of 100% Co spherical precipitates, the model only needs 3 parameters: the Co solubility $C_{eq}$ (i.e. used to calculate the driving force for nucleation), the interfacial energy $\gamma$ and the diffusivity $D$. All 3 parameters are temperature dependent, although the variation of $\gamma$ is expected to be small. To assess the robustness of the model, we used an in-house Matlab®-based implementation and performed a nonlinear least square optimization of the model computed precipitate size and volume fraction to the experimental dataset. The function that was minimized was the following:

$$\chi^2 = \sum_{t,C} \left(\frac{R_{SAXS} - R_{\text{model}}}{\Delta R_{SAXS}}\right)^2 + \sum_{t,C} \left(\frac{f_{v_{SAXS}} - f_{v_{model}}}{\Delta f_{v_{SAXS}}}\right)^2 \qquad (5)$$

where both sums run over all times and all local compositions. $\Delta R_{SAXS}$ and $\Delta f_{v_{SAXS}}$ are the estimated errors on the SAXS determined values of size and volume fraction respectively. The main sources of errors are different for size and volume fraction. Since the SAXS determined size depends only on the shape of the SAXS signal (not on its magnitude), the $\Delta R_{SAXS}$ values depends on the geometrical uncertainties of the experimental setup (dimensions of the detector, sample to detector distance, etc…) which are mostly negligible. The only remaining source of error is the validity of the interpretation model (spherical precipitates, Schulz distribution). We estimate $\Delta R_{SAXS}$ to be 5% (relative to the size). The main source of error for the volume fraction, on the other hand, comes from the intensity scaling and normalization uncertainties[54] which are significantly higher and which we estimate at 15%.

To check for the uniqueness of the solution, several starting parameter sets (within a reasonable range) were tested, each giving similar best-fit parameters. It should be pointed out, however, that this gives no absolute guarantee for the uniqueness of the solution. Each temperature was optimized independently, but for a given temperature, all data points (i.e. the kinetics for each local Co concentration) were considered simultaneously. If each local concentration would have been fitted independently, an even better agreement would have been achieved but with different parameters. A general optimisation guaranties the self-consistency of the obtained set of parameters.

The resulting sizes, volume fractions and number densities are plotted on Figure 8 along with the experimental data. The agreement is generally good, and particularly so for longer ageing times, ie. during the coarsening stages. The comparativediscrepancy between model and experiments at shorter ageing times can possibly be attributed to limitations ofthe classical nucleation theory itself together with assumptions which may not be fully met (such as purely homogeneous nucleation and identical interfacial energies used to describe the Gibbs dividing surface during nucleation and coarsening). Furthermore, while the agreement is excellent for the precipitates size evolution, it is slightly less satisfactory for the volume fractions. This can be at least partly attributed to uncertainties in the absolute scaling of the SAXS data which could be due to uncertainties on the energy of the beam (particularly sensitive close to the Co absorption edge) or other the normalization errors. The SAXS measured sizes on the other hand are insensitive to scaling errors so that the absolute uncertainties on the size values are lower, which may explain the better agreement with the model. Inaccuracies on the volume fractions can also be partly attributed to uncertainties on the local Co concentration, which is the most important input of the precipitation model.

*Table 1 : Parameters obtained by best fit optimization used for the precipitation model*

|       | Co solubility (%) | $\gamma$ (Jm$^{-2}$) | D (m$^2$s$^{-1}$) |
|-------|-------------------|---------------------|-------------------|
| 550°C | 0.194             | 0.250               | 4.30 x 10$^{-18}$ |
| 500°C | 0.192             | 0.197               | 7.62 x 10$^{-19}$ |
| 450°C | 0.199             | 0.184               | 1.66 x 10$^{-20}$ |

The parameters obtained from the best fit optimization of the precipitation model are summarized in Table 1. As expected, the Co diffusivity in the Cu matrix increases with temperature. The literature values[55] for the diffusivities of Co in Cu at 550°C, 500°C and 450°C are 8x10$^{-19}$, 9.5x10$^{-20}$ and 8.4x10$^{-21}$ m$^2$/s, respectively. At 450°C, the value extracted from the precipitation modeling is within a factor of 2 of the literature value, within a factor of 8 at 500°C and within a factor of 5 at 550°C. Considering the uncertainties in diffusivity values this is very good agreement. The interfacial energies extracted from the precipitate modeling (note: the same values are used for modeling both nucleation and coarsening behavior) range between 0.18 and 0.25 Jm$^{-2}$. These compare very well with the values obtained by Stowell (0.18-0.23 Jm$^{-2}$)[41]from analysis of the comprehensive set of nucleation data obtained by Aaronson and LeGoues[40]in the Cu-Co system, and with the value of 0.2 Jm$^{-2}$ reported by Servi and Turnbull in 1966[56]. The Co solubility extracted from the precipitation modelling is found to be fairly constant in the considered temperature range with values around 0.2 at. %. It is to be noted, however, that the absolute value for solubility is very dependent on the accuracy of the local Co concentration (for which there exists some uncertainty) and on the chosen precipitate composition.Using a recent thermodynamic description of the Cu-Co system[57], the Co solubility at 550°C, 500°C and 450°C is calculated to be 0.224%, 0.134% and 0.075% at these respective temperatures.This is a reasonable agreement with the values extracted from the precipitation modeling given the discussed uncertainties on the solubilities. These comparisons illustrate the power of such combinatorial experimentation when combined with modeling. In one set of combinatorial experiments, it has been possible to obtain a sufficiently large dataset as a function of time, temperature and Co alloy content, that application ofrelatively simple precipitation models make it possible to extract values of key precipitate parameters that previously would have taken many years of dedicated studies on separate alloys.

While the concentration of the precipitates could have been allowed to vary during the optimization process, it is likely to lead to non-unique solutions. Additional experimental data on possible temperature dependent precipitate compositions is required to settle the issue.

# 7  Summary

This investigation has demonstrated that it is possible to study precipitation kinetics in-situ in a composition gradient. It has highlighted some of the technical challenges faced when performing such an experiment, particularly on the preparation of the diffusion couples (obtaining a smoothly varying concentration gradient along an adequate distance) and its proper positioning relative to the X-ray beam during the in-situ SAXS experiments. This latter issue could be handled through the design of a dedicated in situ furnace with a wider aperture (e.g. 25 mm) and the preparation of correspondingly wide concentration gradients.

The temporal and spatial resolutions of the time and concentration resolved experiments are sufficient to build a large dataset covering a wide range of supersaturations and ageing times in a single experiment. We performed experiments at 3 temperatures, widening even more the range of precipitation conditions. This dataset was directly compared to the results of a classical class type precipitation model, which, despite its simplicity, was shown to be remarkably robust since it was able to yield realistic values of size and volume fraction for the complete data set with only 3 fitting parameters (solubility, interfacial energy and diffusivity) giving rise to estimated values of these parameters.

This experiment opens a wide field of possible time and concentration experiments enabling a much more robust calibration of phase transformation kinetics models and allowing a much wider set of conditions to be studied simultaneously, hopefully giving rise to better understanding of phase transformations. Experiments similar to those reported here could be performed on similar beamlines containing both SAXS and WAXS detectors opening up the possibility for combinatorial experiments on the competition between precipitation of multiple precipitates which would have important industrial relevance to compositional optimization of existing commercial precipitate strengthened alloys.

# 8 Acknowledgements

The beamtime for this experiment was awarded by the ESRF through experiment MA-1443 and was realized on the French CRG beamline "BM02 – D2AM". The authors would like to thank the D2AM staff for their technical assistance and particularly Dr J.F. Berar for his help concerning the local composition measurement through transmission below and above Co edge. MJS acknowledges the support of CSIRO through the Office of the Chief Executive (OCE) Science Program. CRH gratefully acknowledges the support of the Australian Research Council through the award of a Future Fellowship.

# 9 References

[1] "Materials Genome Initiative for Global Competitiveness, US National Science and Technology Council," 2011.
[2] *The Accelerated Metallurgy Project: www.accmet-project.eu.* .
[3] E. D. Specht, A. Rar, G. M. Pharr, E. P. George, P. Zschack, H. Hong, and J. Ilavsky, "Rapid structural and chemical characterization of ternary phase diagrams using synchrotron radiation," *J. Mater. Res.*, vol. 18, no. 10, pp. 2522–2527, 2003.
[4] J.-C. Zhao, "Reliability of the diffusion-multiple approach for phase diagram mapping," *J. Mater. Sci.*, vol. 39, no. 12, pp. 3913–3925, Jun. 2004.
[5] V. Raghavan, "Addendum ternary and higher order aluminum phase diagram updates," *J. Phase Equilibria Diffus.*, vol. 26, no. 4, pp. 348–348, Oct. 2005.
[6] J.-C. Zhao, "The Diffusion-Multiple Approach to Designing Alloys," *Annu. Rev. Mater. Res.*, vol. 35, no. 1, pp. 51–73, 2005.
[7] J.-C. Zhao, "Combinatorial approaches as effective tools in the study of phase diagrams and composition–structure–property relationships," *Prog. Mater. Sci.*, vol. 51, no. 5, pp. 557–631, Jul. 2006.
[8] C. E. Campbell, W. J. Boettinger, and U. R. Kattner, "Development of a diffusion mobility database for Ni-base superalloys," *Acta Mater.*, vol. 50, no. 4, pp. 775–792, Feb. 2002.
[9] J.-C. Zhao, M. R. Jackson, L. A. Peluso, and L. N. Brewer, "A Diffusion Multiple Approach for the Accelerated Design of Structural Materials," *MRS Bull.*, vol. 27, no. 04, pp. 324–329, Apr. 2002.
[10] S. Vives, P. Bellanger, S. Gorsse, C. Wei, Q. Zhang, and J.-C. Zhao, "Combinatorial Approach Based on Interdiffusion Experiments for the Design of Thermoelectrics:

Application to the Mg-2(Si,Sn) Alloys," *Chem. Mater.*, vol. 26, no. 15, pp. 4334–4337, Aug. 2014.

[11] S. Huxtable, D. G. Cahill, V. Fauconnier, J. O. White, and J.-C. Zhao, "Thermal conductivity imaging at micrometre-scale resolution for combinatorial studies of materials," *Nat. Mater.*, vol. 3, no. 5, pp. 298–301, May 2004.

[12] V. V. Shastry, V. D. Divya, M. A. Azeem, A. Paul, D. Dye, and U. Ramamurty, "Combining indentation and diffusion couple techniques for combinatorial discovery of high temperature shape memory alloys," *Acta Mater.*, vol. 61, no. 15, pp. 5735–5742, Sep. 2013.

[13] S. Ding, Y. Liu, Y. Li, Z. Liu, S. Sohn, F. J. Walker, and J. Schroers, "Combinatorial development of bulk metallic glasses," *Nat. Mater.*, vol. 13, no. 5, pp. 494–500, May 2014.

[14] G. Purdy, J. Ågren, A. Borgenstam, Y. Bréchet, M. Enomoto, T. Furuhara, E. Gamsjager, M. Gouné, M. Hillert, C. Hutchinson, M. Militzer, and H. Zurob, "ALEMI: A Ten-Year History of Discussions of Alloying-Element Interactions with Migrating Interfaces," *Metall. Mater. Trans. A*, vol. 42, no. 12, pp. 3703–3718, Dec. 2011.

[15] M. Gouné, F. Danoix, J. Ågren, Y. Bréchet, C. R. Hutchinson, M. Militzer, G. Purdy, S. van der Zwaag, and H. Zurob, "Overview of the current issues in austenite to ferrite transformation and the role of migrating interfaces therein for low alloyed steels," *Mater. Sci. Eng. R Rep.*, vol. 92, pp. 1–38, Jun. 2015.

[16] A. Borgenstam and M. Hillert, "Massive transformation in the Fe–Ni system," *Acta Mater.*, vol. 48, no. 11, pp. 2765–2775, Jun. 2000.

[17] B. Chehab, "Etude de la forgeabilite d'un acier inoxydable ferritique biphase a chaud," Thèse de l'Institut National Polytechnique de Grenoble, 2007.

[18] A. Borgenstam and J. M. Ericsson, "Determination of the critical carbon content for growth of acicular ferrite," in *International Conference on Solid-Solid Phase Transformations in Inorganic Materials (PTM 2005)*, Phoenix, USA, 2005, pp. 105–110.

[19] A. Phillion, H. W. Zurob, C. R. Hutchinson, H. Guo, D. V. Malakhov, J. Nakano, and G. R. Purdy, "Studies of the influence of alloying elements on the growth of ferrite from austenite under decarburization conditions: Fe-C-Nl alloys," *Metall. Mater. Trans. A*, vol. 35, no. 4, pp. 1237–1242, Apr. 2004.

[20] C. R. Hutchinson, A. Fuchsmann, H. S. Zurob, and Y. Brechet, "A novel experimental approach to identifying kinetic transitions in solid state phase transformations," *Scr. Mater.*, vol. 50, no. 2, pp. 285–290, 2004.

[21] C. W. Sinclair, C. R. Hutchinson, and Y. Bréchet, "The Effect of Nb on the Recrystallization and Grain Growth of Ultra-High-Purity α-Fe: A Combinatorial Approach," *Metall. Mater. Trans. A*, vol. 38, no. 4, pp. 821–830, Jun. 2007.

[22] T. Miyazaki, "Development of 'Macroscopic Composition Gradient Method' and its application to the phase transformation," *Prog. Mater. Sci.*, vol. 57, no. 6, pp. 1010–1060, Jul. 2012.

[23] S. Hata, K. Shiraishi, M. Itakura, N. Kuwano, T. Nakano, and Y. Umakoshi, "Long-period ordering in a TiAl single crystal with a gradient composition," *Philos. Mag. Lett.*, vol. 85, no. 4, pp. 175–185, Apr. 2005.

[24] E. Contreras-Piedras, H. J. Dorantes-Rosales, V. M. López-Hirata, F. Hernández Santiago, J. L. González-Velázquez, and F. I. López-Monrroy, "Analysis of precipitation in Fe-rich Fe–Ni–Al alloys by diffusion couples," *Mater. Sci. Eng. A*, vol. 558, pp. 366–370, Dec. 2012.

[25] R. K. W. Marceau, C. Qiu, S. P. Ringer, and C. R. Hutchinson, "A study of the composition dependence of the rapid hardening phenomenon in Al–Cu–Mg alloys using diffusion couples," *Mater. Sci. Eng. A*, vol. 546, pp. 153–161, Jun. 2012.

[26] V. Gerold and G. Kostorz, "Small-angle scattering applications to materials science," *J. Appl. Crystallogr.*, vol. 11, no. 5, pp. 376–404, Oct. 1978.

[27] G. Kostorz, "Small-angle scattering studies of phase separation and defects in inorganic materials," *J. Appl. Crystallogr.*, vol. 24, no. 5, pp. 444–456, Oct. 1991.

[28] P. Fratzl, "Small-angle scattering in materials science - a short review of applications in alloys, ceramics and composite materials," *J. Appl. Crystallogr.*, vol. 36, no. 3, pp. 397–404, Apr. 2003.

[29] F. De Geuser and A. Deschamps, "Precipitate characterisation in metallic systems by small-angle X-ray or neutron scattering," *Comptes Rendus Phys.*, vol. 13, no. 3, pp. 246–256, Apr. 2012.

[30] A. Deschamps and F. De Geuser, "Quantitative Characterization of Precipitate Microstructures in Metallic Alloys Using Small-Angle Scattering," *Metall. Mater. Trans. A*, vol. 44, no. 1, pp. 77–86, Jan. 2013.

[31] M. Dumont, A. Steuwer, A. Deschamps, M. Peel, and P. J. Withers, "Microstructure mapping in friction stir welds of 7449 aluminium alloy using SAXS," *Acta Mater*, vol. 54, no. 18, pp. 4793–4801, 2006.

[32] A. Steuwer, M. Dumont, J. Altenkirch, S. Birosca, A. Deschamps, P. B. Prangnell, and P. J. Withers, "A combined approach to microstructure mapping of an Al–Li AA2199 friction stir weld," *Acta Mater.*, vol. 59, no. 8, pp. 3002–3011, May 2011.

[33] F. De Geuser, B. Malard, and A. Deschamps, "Microstructure mapping of a friction stir welded AA2050 Al–Li–Cu in the T8 state," *Philos. Mag.*, vol. 94, no. 13, pp. 1451–1462, 2014.

[34] G. Fribourg, Y. Brechet, A. Deschamps, and A. Simar, "Microstructure-based modelling of isotropic and kinematic strain hardening in a precipitation-hardened aluminium alloy," *Acta Mater.*, vol. 59, no. 9, pp. 3621–3635, May 2011.

[35] P. Schloth, J. N. Wagner, J. L. Fife, A. Menzel, J.-M. Drezet, and H. V. Swygenhoven, "Early precipitation during cooling of an Al-Zn-Mg-Cu alloy revealed by in situ small angle X-ray scattering," *Appl. Phys. Lett.*, vol. 105, no. 10, p. 101908, Sep. 2014.

[36] H. Wendt and P. Haasen, "Atom probe field ion microscopy of the decomposition of Cu-2.7 at% Co," *Scr. Metall.*, vol. 19, no. 9, pp. 1053–1058, 1985.

[37] M. Takeda, H. Yamada, S. Yoshida, K. Shimasue, T. Endo, and J. Van Landuyt, "TEM study and Monte-Carlo simulation of nano-scale Co particles precipitated in a Cu matrix," *Phys. Status Solidi A*, vol. 198, no. 2, pp. 436–442, Aug. 2003.

[38] M. Takeda, K. Inukai, N. Suzuki, G. Shinohara, and H. Hashimoto, "Precipitation Behaviour of Cu☐Co Alloys," *Phys. Status Solidi A*, vol. 158, no. 1, pp. 39–46, 1996.

[39] R. P. Setna, J. M. Hyde, A. Cerezo, G. D. W. Smith, and M. F. Chisholm, "Position sensitive atom probe study of the decomposition of a Cu-2.6at%Co alloy," *Appl. Surf. Sci.*, vol. 67, no. 1–4, pp. 368–379, Apr. 1993.

[40] H. I. Aaronson and F. K. LeGoues, "An assessment of studies on homogeneous diffusional nucleation kinetics in binary metallic alloys," *Metall. Trans. A*, vol. 23, no. 7, pp. 1915–1945, 1992.

[41] M. J. Stowell, "Precipitate nucleation: does capillarity theory work?," *Mater. Sci. Technol.*, vol. 18, no. 2, pp. 139–144, 2002.

[42] R. Wagner, R. Kampmann, and P. W. Voorhees, "Homogeneous Second-Phase Precipitation," in *Phase Transformations in Materials*, Wiley-VCH Verlag GmbH & Co., 2005, pp. 309–407.

[43] M. Perez, M. Dumont, and D. Acevedo-Reyes, "Implementation of classical nucleation and growth theories for precipitation," *Acta Mater*, vol. 56, no. 9, pp. 2119–2132, 2008.

[44] J. D. Robson, "A new model for prediction of dispersoid precipitation in aluminium alloys containing zirconium and scandium," *Acta Mater.*, vol. 52, pp. 1409–1421, 2004.

[45] M. Nicolas and A. Deschamps, "Characterisation and modelling of precipitate evolution in an Al-Zn-Mg alloy during non-isothermal heat treatments," *Acta Mater.*, vol. 51, no. 20, pp. 6077–6094, 2003.

[46] A. Deschamps, C. Sigli, T. Mourey, F. de Geuser, W. Lefebvre, and B. Davo, "Experimental and modelling assessment of precipitation kinetics in an Al-Li-Mg alloy," *Acta Mater.*, vol. 60, no. 5, pp. 1917–1928, Mar. 2012.

[47] O. Glatter and O. Kratky, Eds., *Small-Angle X-ray Scattering*. London: Academic Press, 1982.

[48] B. J. Heuser, "Small-angle neutron scattering study of dislocations in deformed single-crystal copper," *J. Appl. Crystallogr.*, vol. 27, no. 6, pp. 1020–1029, 1994.

[49] J. S. Pedersen, "Determination of size distribution from small-angle scattering data for systems with effective hard-sphere interactions," *J. Appl. Crystallogr.*, vol. 27, no. 4, pp. 595–608, 1994.

[50] G. V. Schulz, "The molecular weight distribution of oligomers," *Polymer*, vol. 23, no. 4, pp. 497–498, Apr. 1982.

[51] G. M. Novotny and A. J. Ardell, "Precipitation of Al 3 Sc in binary Al–Sc alloys," *Mater. Sci. Eng. A*, vol. 318, no. 1, pp. 144–154, 2001.

[52] M. Hättestrand and H.-O. Andrén, "Evaluation of particle size distributions of precipitates in a 9% chromium steel using energy filtered transmission electron microscopy," *Micron*, vol. 32, no. 8, pp. 789–797, 2001.

[53] S. R. Aragón and R. Pecora, "Theory of dynamic light scattering from polydisperse systems," *J. Chem. Phys.*, vol. 64, no. 6, pp. 2395–2404, Mar. 1976.

[54] F. Zhang, J. Ilavsky, G. Long, J. Quintana, A. Allen, and P. Jemian, "Glassy Carbon as an Absolute Intensity Calibration Standard for Small-Angle Scattering," *Metall. Mater. Trans. A*, vol. 41, no. 5, pp. 1151–1158, May 2010.

[55] H. P. J. Wijn, Ed., *Landolt-Börnstein - Group III Condensed Matter. Alloys and Compounds of d-Elements with Main Group Elements. Part 1*, vol. 32B. Berlin/Heidelberg: Springer-Verlag, 1999.

[56] I. S. Servi and D. Turnbull, "Thermodynamics and kinetics of precipitation in the copper-cobalt system," *Acta Metall.*, vol. 14, no. 2, pp. 161–169, Feb. 1966.

[57] M. Palumbo, S. Curiotto, and L. Battezzati, "Thermodynamic analysis of the stable and metastable Co–Cu and Co–Cu–Fe phase diagrams," *Calphad*, vol. 30, no. 2, pp. 171–178, Jun. 2006.

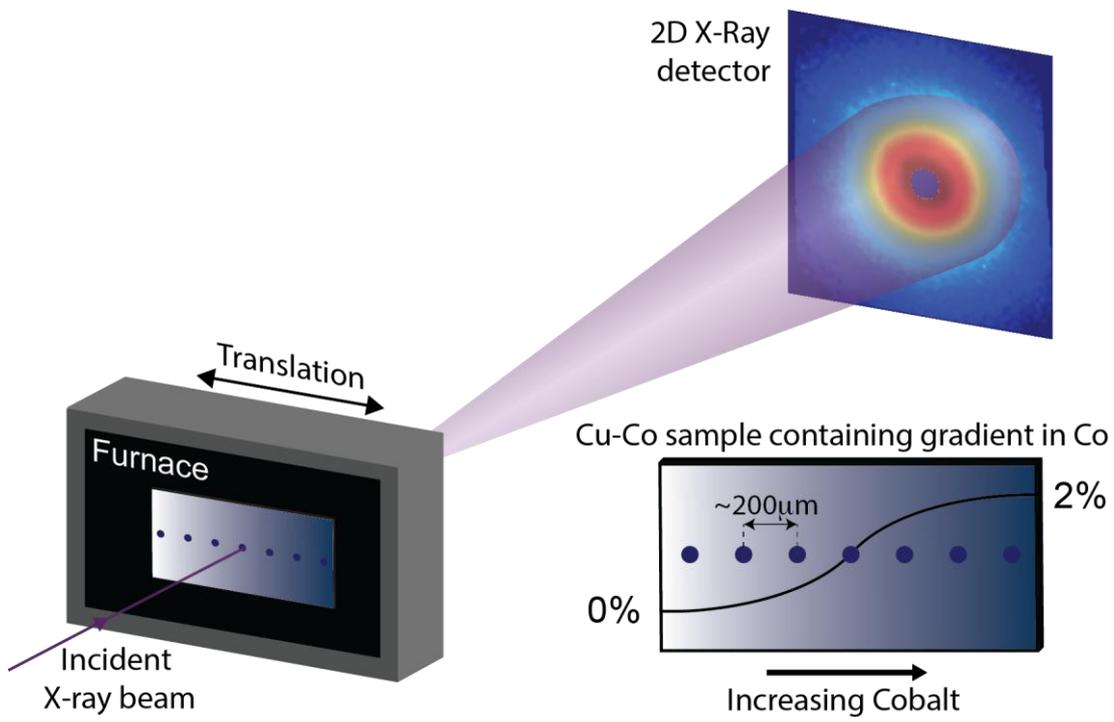

*Figure 1 : Schematic representation of the experimental set-up showing the scanning of the incident x-ray beam across the composition gradient during elevated temperature aging treatment*

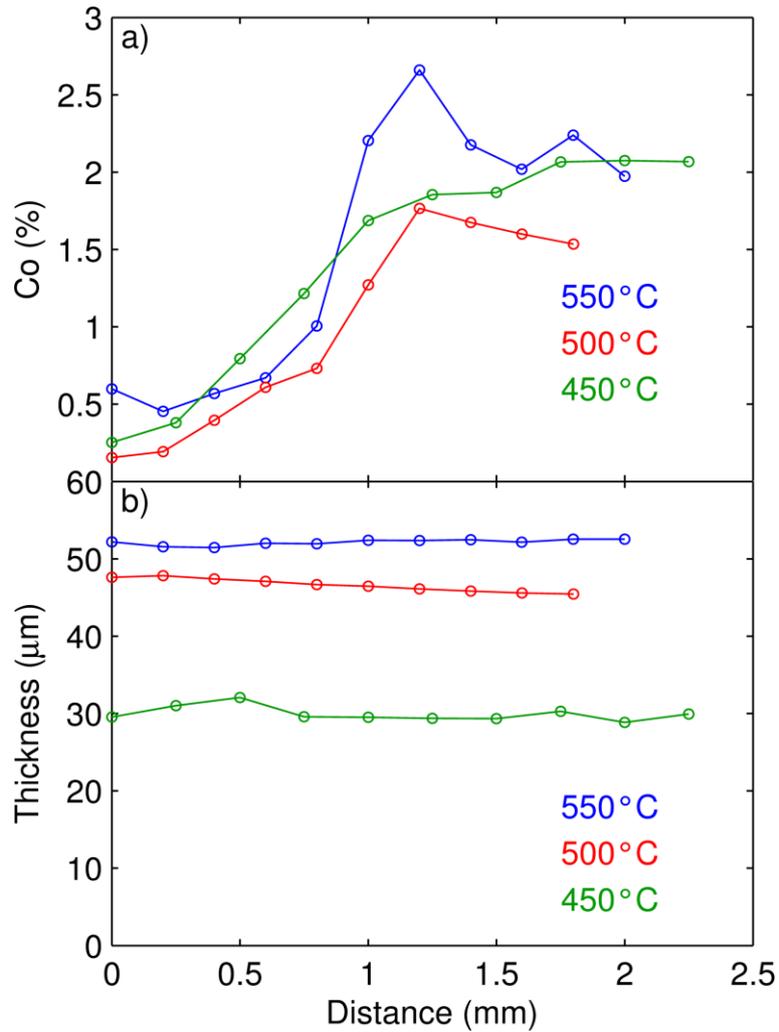

*Figure 2 : a) Composition profiles and b) local sample thicknesses across the diffusion couples determined based on the intensity of the transmitted X-ray beam during the in situ SAXS experiment.*

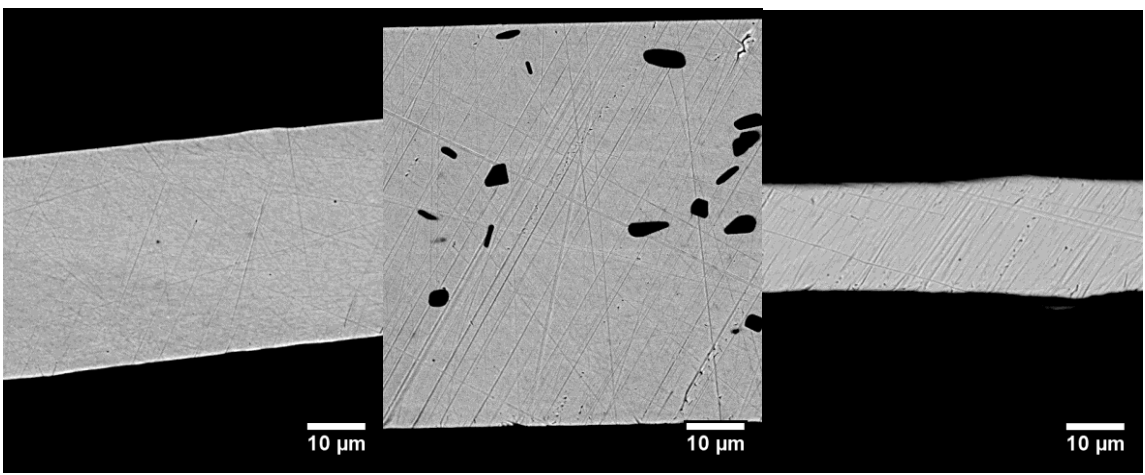

*Figure 3 : BSE micrographs collected from a) the pure Cu end, b) diffusion zone and c) Co rich end of a solution treated and quenched diffusion couple. A number of dark spots (< 1 μm) are visible at the Co rich end of the sample (c), and are thought to be small voids. Spot analyses of the large (~5 μm) inclusions found in the diffusion zone (b) indicate that they are Co oxides.*

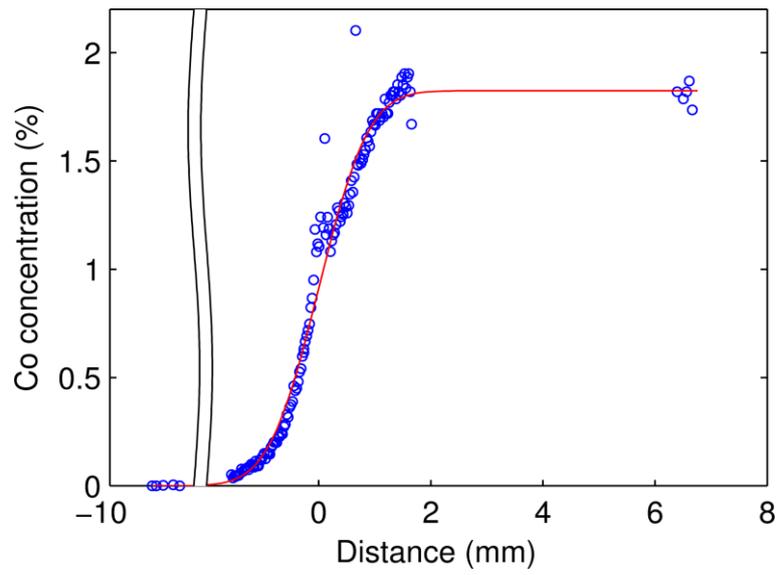

*Figure 4 : Co concentration profile measured by EPMA over the length of the solution treated diffusion couple. The average composition at the Co-rich end of the diffusion couple is 1.82 at.%. A model of the diffusion profile is also shown in red.*

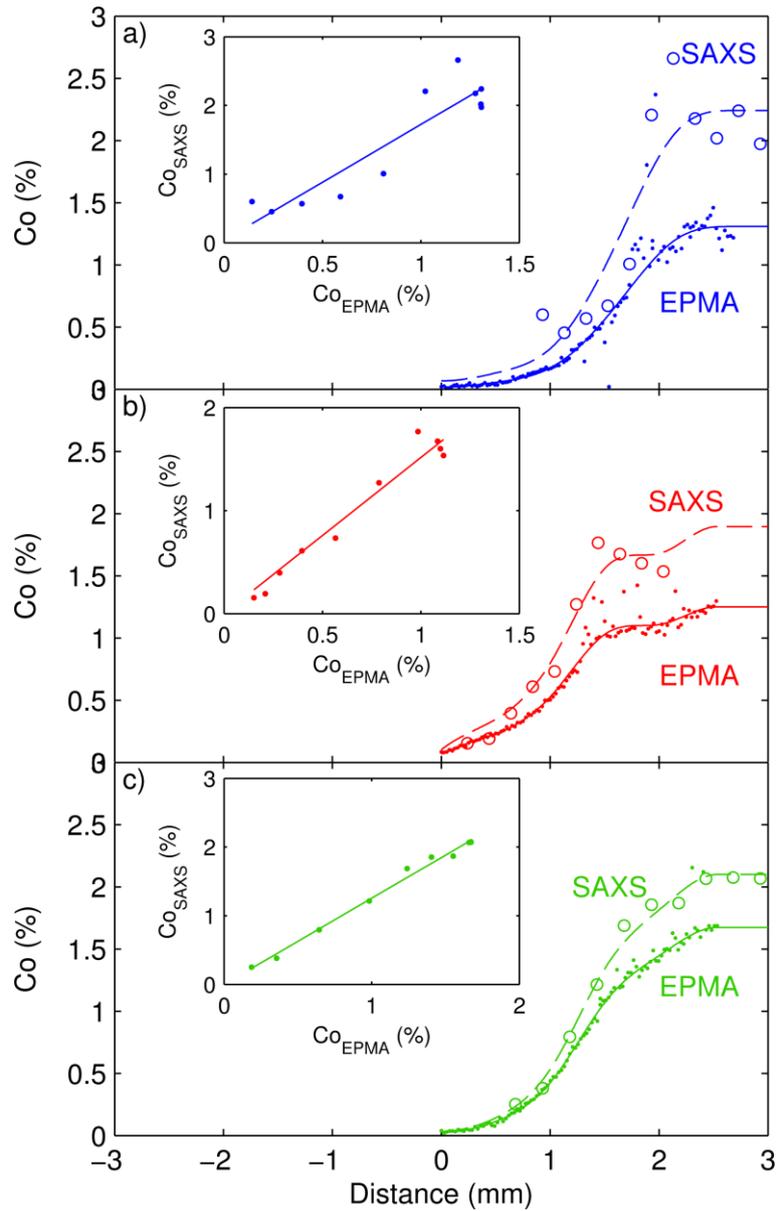

*Figure 5 : Comparison of the Co concentration profiles measured using EPMA and SAXS for the samples aged in situ at a) 550°C, b) 500°C and c) 450°C. The insets show the linear relationship between EPMA and SAXS results. The dashed trend line used for SAXS is the linearly transformed EPMA trend line.*

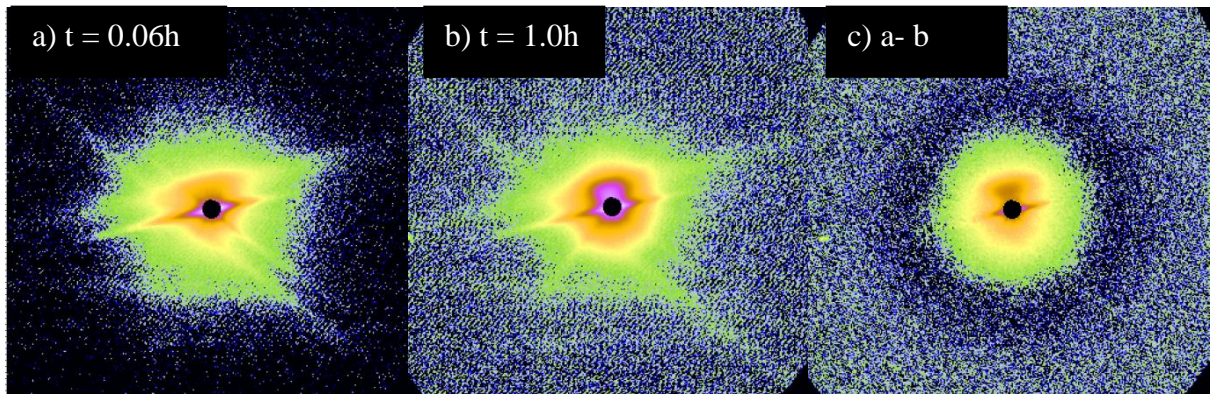

*Figure 6 : SAXS image collected from the middle of the diffusion zone of the 550°C sample. a) initial image (t=0.06h) and b) image collected after 1h of ageing. Note the multiple intense streaks passing through or near the beamstop. c) Image after 1h minus initial image. While some streak features still appear, the signal is now mostly isotropic, as expected from spherical precipitates. The same colour scale is used for all the images.*

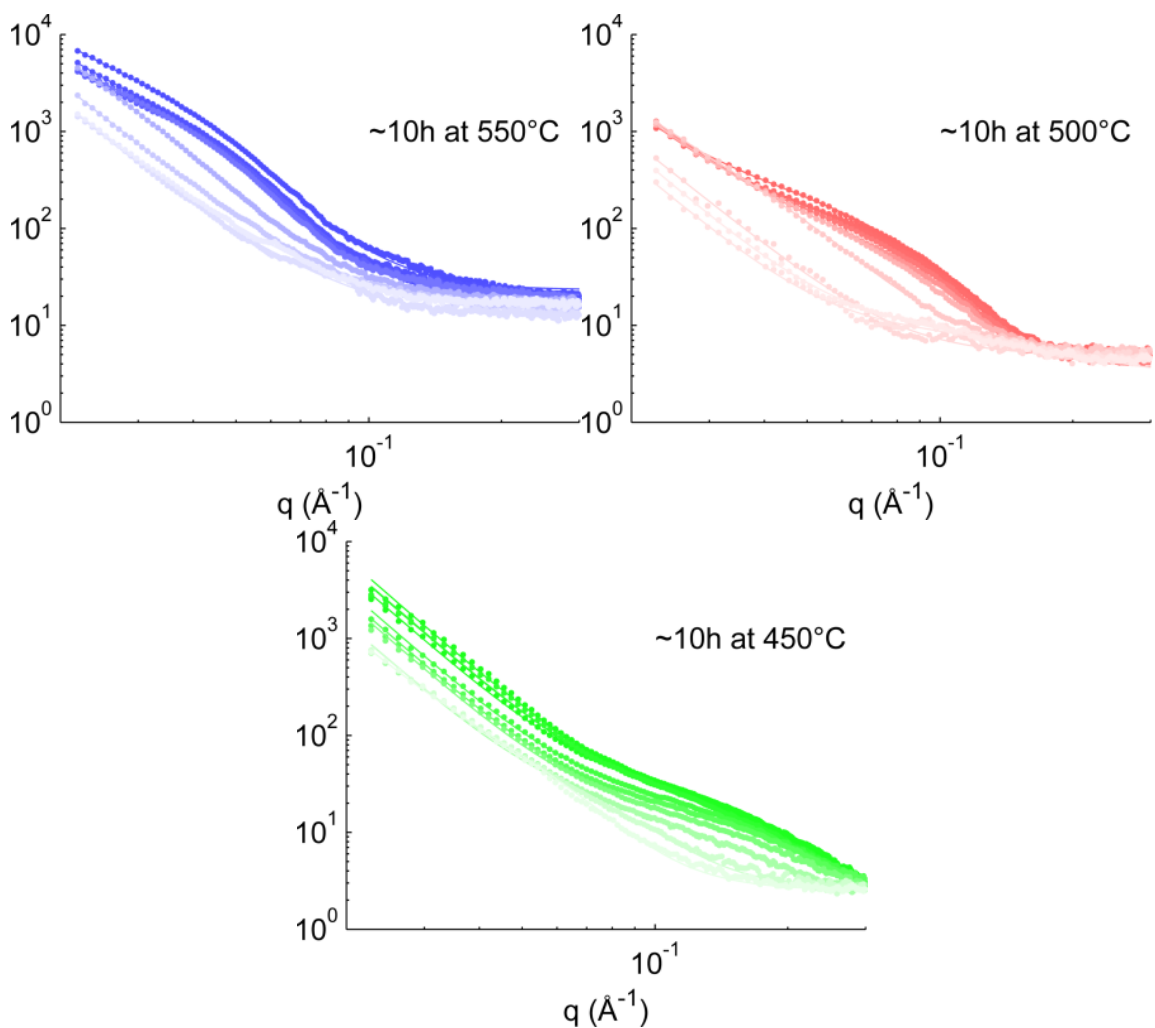

*Figure 7 : experimental SAXS patterns obtained after 10h at a) 550°C, b) 500°C and c) 450°C. The color shade isrelated to the local Co content (darker = more Co). The lines are the best fits according to Eq. (4).*

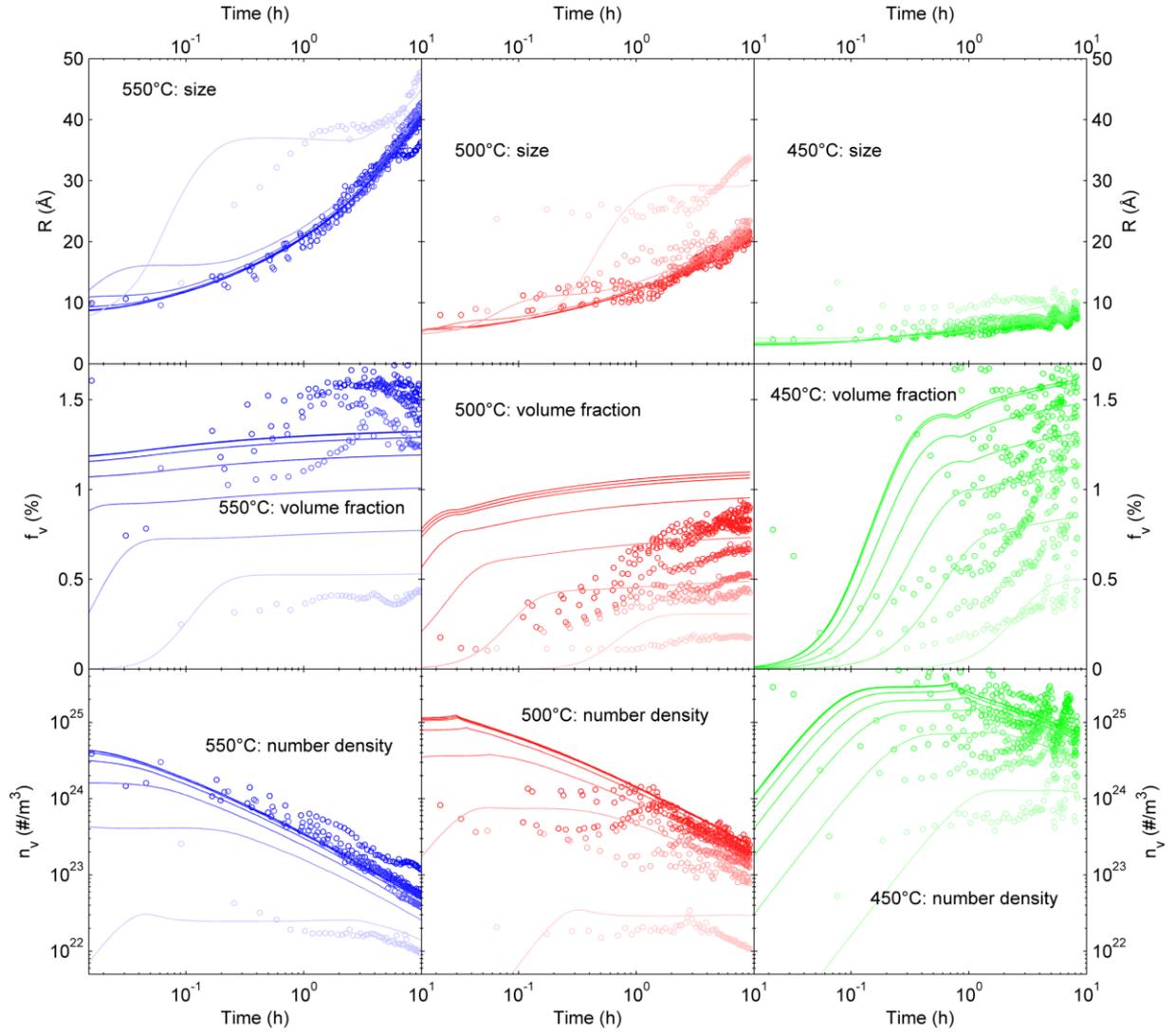

*Figure 8: Time and concentration resolved precipitation kinetics of the Cu-(CuCo) diffusion couples obtained for 3 temperatures, 550°C (blue), 500°C (red) and 450°C (green). The circles are the SAXS experimental results, the lines are the results of the precipitation model. The color shades are related to the local Co content (darker=more Co). For easier reference to the precise local composition kinetics, each individual precipitation kinetics has been plotted separately in the supplementary materials.*

*Table 1 : Parameters obtained by best fit optimization used for the precipitation model*

|  | Co solubility (%) | $\gamma$ (Jm$^{-2}$) | D (m$^2$s$^{-1}$) |
|---|---|---|---|
| 550°C | 0.194 | 0.250 | 4.30 x 10$^{-18}$ |
| 500°C | 0.192 | 0.197 | 7.62 x 10$^{-19}$ |
| 450°C | 0.199 | 0.184 | 1.66 x 10$^{-20}$ |